\documentclass[preprint,aps,draft]{revtex4}

\usepackage{graphicx}
\usepackage{dcolumn}
\usepackage{bm}
\usepackage{times}
\usepackage{textcomp}
\usepackage{amsmath}

\begin{document}
\preprint{APS/123-QED}

\title{Mode conversion by symmetry breaking of propagating spin waves}

\author{P.~Clausen,$^1$ K.~Vogt,$^{1,2}$ H.~Schultheiss,$^{1,3}$ S. Sch\"afer,$^{1,4}$ B.~Obry,$^1$  G.~Wolf,$^1$ P.~Pirro,$^1$ B.~Leven,$^1$ and B. Hillebrands$^1$}
\affiliation{$^1$Fachbereich Physik and Forschungszentrum OPTIMAS, Technische Universit\"at Kaiserslautern, D-67663 Kaiserslautern, Germany\\
$^2$Graduate School of Excellence Material Science in Mainz, Staudinger Weg 9, D-55128 Mainz, Germany\\
$^3$Material Science Division, Argonne National Laboratory, USA\\
$^4$MINT Center, University of Alabama, Tuscaloosa, AL 35487, USA}

\date{July 22, 2011}

\begin{abstract}
We study spin-wave transport in a microstructured Ni$_{81}$Fe$_{19}$ wave\-guide exhibiting broken translational symmetry. We observe the conversion of a beam profile composed of symmetric spin-wave width modes with odd numbers of antinodes $n=1,3,\ldots$ into a mixed set of symmetric and asymmetric modes. Due to the spatial homogeneity of the exciting field along the used mic\-ro\-strip antenna, quantized spin-wave modes with an even number $n$ of antinodes across the stripe's width cannot be directly excited. We show that a break in translational symmetry may result in a partial conversion of even spin-wave waveguide modes.
\end{abstract}


\maketitle

For the advancement of the emergent field of magnonics and magnon-spintronics~\cite{Khitun2010, Kruglyak2010, Serga2010, Kruglyak2006, Schneider2005}, the understanding and characterization of spin waves in magnetic microstructures are essential. Many mechanisms for the excitation of spin waves on the micrometer length scale either by uniform microwave field~\cite{Au2011}, mic\-ro\-strip antennas~\cite{Vogt2009, Demidov2009b}, spin-transfer torque~\cite{Slonczewski1996,Demidov2011}, or oscillating domain walls~\cite{Hermsdoerfer2009} have been explored recently. However, the transport of spin waves in wave\-guide structures was mainly focused on one-dimensional geometries~\cite{Bauer1997,Demidov2009c}. The implementation of spin waves for the transport and processing of information even in devices as simple as a signal splitter will require changes of the spin-wave propagation direction. Modifications in the shape of a magnetic wave\-guide often result in inhomogeneous demagnetizing fields and non-uniform magnetization distributions. However, the spin-wave dispersion as well as the mode profiles adiabatically adjust to the local variations in the internal field~\cite{Bayer2004}. This makes the overall propagation and quantization pattern complex and difficult to predict. 

\begin{figure}
  \includegraphics[width=0.9\columnwidth]{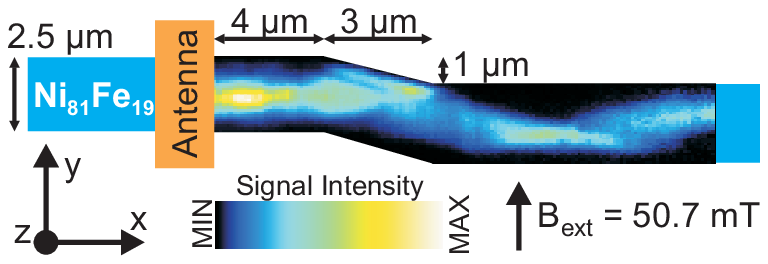}
  \caption{\label{Fig1} (Color online) Concept of the investigated lateral spin-wave wave\-guide. The distribution of the spin-wave intensity measured by Brillouin light scattering (BLS) microscopy is color coded where black (white) indicates minimum (maximum) intensity, respectively. The measured intensity has been normalized to compensate for damping.}
\end{figure}

A way out of this problem are specific design rules for the layout of spin-wave wave\-guide structures. In this Letter, we demonstrate that such design rules exist and specific mode properties can be utilized. We demonstrate this using the example of spin-wave transport in a wave\-guide with broken translational symmetry. The wave\-guide is shifted transverse to the spin-wave propagation direction resulting in a parallel offset of the wave\-guide axis behind the skew section (see Fig.~\ref{Fig1}). Despite being only a rather small deviation from a straight stripe, we show, that this skew has a profound impact on the spin-wave modes propagating in the wave\-guide.

The schematic layout of the investigated microstructure is shown in Fig.~\ref{Fig1}. A spin-wave wave\-guide with dimensions of $w=2.5$\,$\mu$m  $\times$ $l=100$\,$\mu$m is fabricated from a 60\,nm thick Ni$_{81}$Fe$_{19}$ film using electron-beam lithography and conventional lift-off techniques. Spin waves are excited by the magnetic Oersted field of a 2\,$\mu$m wide and 500\,nm thick Cu mic\-ro\-strip antenna driven with frequencies $\omega_{\mathrm{rf}}/2\pi$ up to 18\,GHz. In a distance of 4\,$\mu$m from the antenna, a 3\,$\mu$m long skew results in a 1\,$\mu$m parallel offset of the wave\-guide. We apply a static magnetic field of $B_\mathrm{ext}=50.7\,\mathrm{mT}$ along the short axis of the wave\-guide ($y$-direction) to ensure the excitation of propagating spin waves travelling perpendicular to the magnetization direction with high group velocities. This allows for the investigation of propagation phenomena over large distances by means of Brillouin light scattering (BLS) microscopy~\cite{Demidov2004, Vogt2009} as can be seen in the spin-wave intensity pattern overlaid in Fig.~\ref{Fig1}.

\begin{figure}
  \includegraphics[width=0.9\columnwidth]{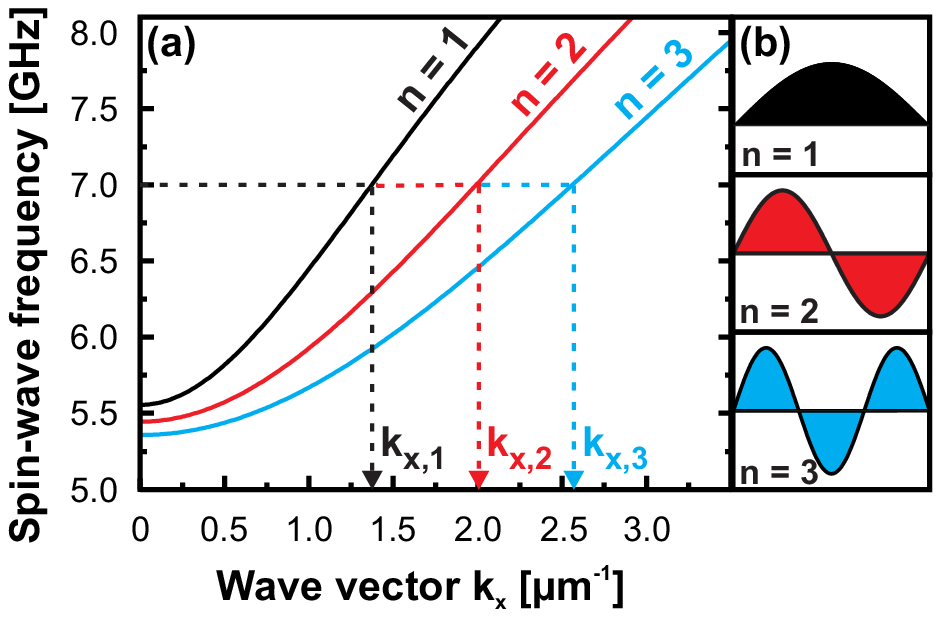}
	\caption{\label{Fig2} (Color online) (a) Frequencies of different lateral spin-wave modes as a function of the wave vector component along the wave\-guide ($x$-direction). The dashed lines depict the wave vectors for a fixed excitation frequency. (b) Schematic spin-wave mode profiles in $y$-direction.}
\end{figure}

A wave\-guide with a finite transversal width causes a discretization of the spin-wave wave vector in $y$-direction leading to a well-defined set of dispersion relations as shown in Fig.~\ref{Fig2}(a). These dispersion relations display the frequency dependence of the spin-wave modes as a function of the wave vector component $k_{x}$ in $x$-direction along the wave\-guide. The wave vector in $y$-direction, perpendicular to the wave\-guide, is quantized according to $k_{n,y}={n}\pi/w_\mathrm{eff}$, where the mode number $n$ equals the number of antinodes in $y$-direction across the width of the stripe. In this geometry, the demagnetizing fields lead to a reduction of the effective quantization width $w_\mathrm{eff}$ smaller than the geometrical width $w$ of the stripe. The calculation of the dispersion relations is made following the approach in~\cite{Vogt2009} for an internal magnetic field of $B_\mathrm{int}=39.5\,\mathrm{mT}$ and standard material parameters for Ni$_{81}$Fe$_{19}$ as summarized in~\cite{NiFe}. The internal magnetic field is extracted from a micromagnetic simulation (oommf code, see~\cite{Donahue1999}) and set to an averaged value over the assumed quantization width. The smaller value compared to the externally applied field is due to the demagnetizing fields originating from the magnetic charges at the boundaries of the wave\-guide. 

In the further discussion, we assume a sinusoidal mode profile in $y$-direction, as schematically displayed in Fig.~\ref{Fig2}(b) for the three lowest width modes. This leads to an overall spatial spin-wave amplitude given by
	\begin{equation}\label{eq1}
	{\psi_{n}(x, y, t)} = A_{n} e^{-i(k_{x}x-\omega_{\mathrm{rf}}t)}\cdot \sin\Bigl(\frac{n\pi}{w}y\Bigr),
	\end{equation}
where $A_{n}$ is the maximum amplitude of the respective partial mode. For a given excitation frequency (7\,GHz in Fig.~\ref{Fig2}(a)) several modes with different $n$ can be excited simultaneously. These modes interfere with each other and form a stationary intensity distribution~\cite{Buttner1998,Demidov2008}. The resulting time-averaged interference pattern is the coherent superposition 
	\begin{equation}\label{eq2}
	I(x,y)=\Bigl| \sum_{n}{\psi_{n}(x, y)}\Bigr|^{2}
	\end{equation}
of the amplitudes given in equation~(\ref{eq1}) and reflects the symmetry of the interfering modes. Only in the case when all spin-wave modes have an odd mode number $n$, the resulting intensity distribution is symmetric with respect to the center of the wave\-guide. 

Since in our experiment the exciting field of the antenna is homogeneous across the wave\-guide's width, only spin-wave modes with a net component of the dynamic magnetization, i.e., odd modes with $n=1,3,5,\ldots$, can be directly excited. Therefore, only a symmetric spin-wave interference pattern is found in the first 4\,$\mu$m behind the antenna as can be seen in the measured BLS intensity in Fig.~\ref{Fig1}.

Inside the skew section, the spin-wave intensity distribution changes significantly. The incident spin-wave beam is not simply reflected at the edge of the skew section but rather attenuated. Furthermore, it excites a different mode which is strongly localized at the edge of the wave\-guide in a spin-wave well~\cite{Jorzick2002}. 
Calculations show, that the inhomogeneity of the internal magnetic field $B_\mathrm{int}$, which is modified by the demagnetizing field of the stripe, traps the spin waves in a region extending only several hundred nanometers from the edge of the stripe which is comparable to the lateral dimensions of the edge mode in Fig.~\ref{Fig1}.

Behind the skew section, in 7\,$\mu$m distance from the antenna, the interference pattern has an asymmetric component. This asymmetric spin-wave intensity distribution observed behind the skew can be described in good approximation as a superposition of the lowest even and odd mode with $n=1$ and $n=2$. This conversion from a purely symmetric system to a mixed set of symmetric and asymmetric spin-wave modes is caused by an asymmetric distribution of the dynamic magnetization, which originates from the originally excited modes ending up with their intensity maximum displaced from the center of the stripe by the skew. As a result, the asymmetric mode with $n=2$ is excited.

To quantitatively analyze this effect we consider the periodicity of the spin-wave intensity. For each stationary spin-wave intensity distribution, being either symmetric or asymmetric, its periodicity $d_\text{per}$ in $x$-direction is determined by the difference $\Delta k_{x}(\omega_{\mathrm{rf}})$ of the wave vector components $k_{x}$ of the interfering spin waves and is a function of the excitation frequency $\omega_{\mathrm{rf}}$. In case the two lowest spin-wave modes are excited, this leads to a periodicity $d_\text{per}$ given by:
\begin{equation}\label{eq3}
	  d_\text{per}(\omega_{\mathrm{rf}}) = \frac{2\pi}{\bigl({{k}_{x,1}(\omega_{\mathrm{rf}}) - {k}_{x,2}}(\omega_{\mathrm{rf}})\bigr)}.
\end{equation}
In the experiment, the values of $k_{x,1}$ and $k_{x,2}$ are determined by the spin-wave dispersion relations shown in Fig.~\ref{Fig2}(a). Consequently, the periodicity $d_\text{per}$ can be controlled by either tuning the excitation frequency $\omega_{\mathrm{rf}}$ or the externally applied magnetic field $B_\mathrm{ext}$. 

To support the stated interpretation of the asymmetric interference pattern being composed of the spin-wave modes with $n=1$ and $n=2$, we measured and calculated the intensity distributions of these two interfering modes for excitation frequencies $\omega_{\mathrm{rf}}/2\pi$ ranging from 6.6 to 8.0\,GHz. For calculating the interference pattern we use equations~(\ref{eq1}) and (\ref{eq2}) with $n=1,2$ and the wave vectors determined from the corresponding dispersion relations as in Fig.~\ref{Fig2}(a) for each excitation frequency. The amplitudes $A_{n}$ are free parameters in the calculation and were set to match the measured intensity distributions. However, their specific values will not influence the periodicity of the interference pattern.

\begin{figure}
	\includegraphics[width=0.9\columnwidth]{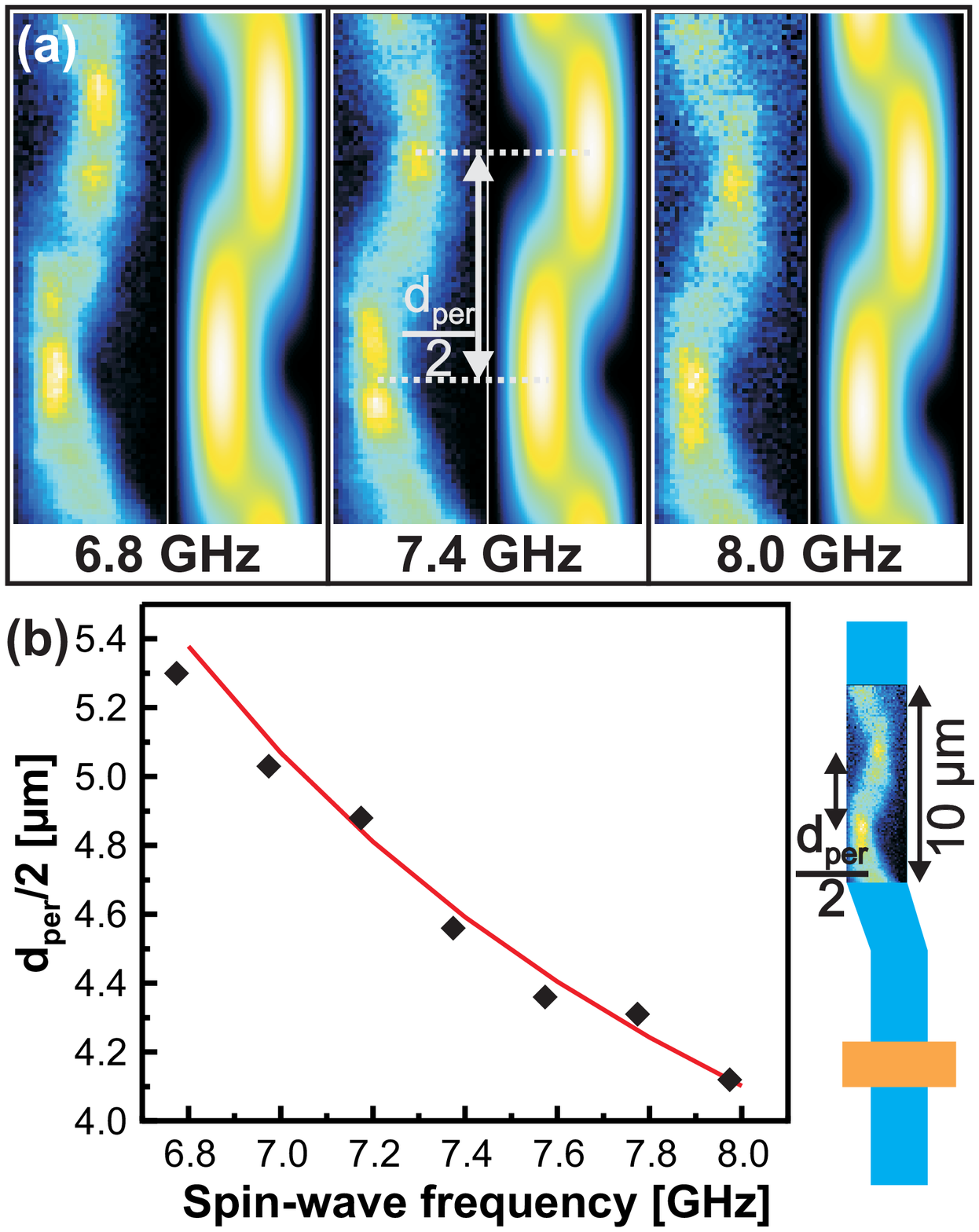}
	\caption{\label{Fig3} (Color online) (a) Comparison of the measured (left sub-panels) and calculated (right sub-panels) spin-wave interference patterns for different frequencies. (b) Black diamonds show the periodicity $d_\text{per}$ extracted from the interference patterns as illustrated in (a) as a function of the spin-wave frequency. The curve depicts calculated values.}
\end{figure}

For three exemplary excitation frequencies, the intensity distributions measured behind the skew can be seen in Fig.~\ref{Fig3}(a) both for experiment (left sub-panels) and calculations (right sub-panels). The resulting change in the periodicity $d_\text{per}$ of the snake-like mode pattern is evident in the experiment as well as in the calculations, both showing an excellent agreement.

Figure~\ref{Fig3}(b) summarizes the results obtained from BLS microscopy measurements (filled symbols) in direct comparison with the theoretical expectations (line), which are given by equation~(\ref{eq3}).  For increasing excitation frequencies, the split in the dispersion relations becomes more pronounced leading to an increase of $\Delta k_{x}(\omega_{\mathrm{rf}})$. Therefore, the periodicity $d_\text{per}$ decreases in agreement with equation~(\ref{eq3}) as can be seen in Fig.~\ref{Fig3}(a) and (b). The agreement between measured data and calculations further supports our assumption of observing interference between the modes $n=1$ and $n=2$ behind the skew.

In conclusion, we were able to manipulate the spin-wave mode patterns of propagating spin waves in a microscopic wave\-guide from being a superposition of symmetric modes to a mixed system of symmetric/asymmetric spin-wave modes. This controlled mode conversion is achieved by a shift of the wave\-guide transverse to the propagation direction of the spin waves creating a variation in the magnetization distribution within the skew section and leading to a strong concentration of the spin-wave amplitude on one side of the wave\-guide. Ultimately, this inhomogeneity causes the excitation of asymmetric spin-wave modes behind the skew section. This is not only an eminent chance to create spin-wave modes in microstructures which cannot be excited by antennas alone due to their homogeneous field. The study of spin-wave propagation and mode conversion in this structure may also lead to further insight into the interaction between spin waves and their excitation via the local dynamic magnetization. In addition, it is a first step to better understand and utilize spin-wave transport  in truly two-dimensional microstructures, an essential step for the implementation of magnonics and magnon-spintronics into future devices.

The authors thank Dr.~P.~A.~Beck for deposition of the magnetic thin film and the Nano$+$Bio Center of the Technische Universit\"at Kaiserslautern for assistance in sample preparation. Financial support by the Carl-Zeiss-Stiftung and the Graduiertenkolleg 792 is gratefully acknowledged.


\begin{thebibliography}{1}

\bibitem{Khitun2010}A. Khitun, M. Bao, and K.~L. Wang, J. Phys. D \textbf{43}, 264005 (2010).

\bibitem{Kruglyak2010}V.~V. Kruglyak, S.~O. Demokritov, and D. Grundler, J. Phys. D \textbf{43}, 264001 (2010).

\bibitem{Serga2010}A.~A. Serga, A.~V. Chumak, and B. Hillebrands, J. Phys. D \textbf{43}, 264002 (2010).

\bibitem{Kruglyak2006}V.~V. Kruglyak and R.~J. Hicken, J. Magn. Magn. Mater. \textbf{306}, 191 (2006).

\bibitem{Schneider2005}T. Schneider, A.~A. Serga, B. Leven, B. Hillebrands, R.~L. Stamps, and M.~P. Kostylev, Appl. Phys. Lett. \textbf{92}, 022505 (2008).

\bibitem{Au2011}Y. Au, T. Davison, E. Ahmad, P.~S. Keatley, R.~J. Hicken, and V.~V. Kruglyak, Appl. Phys. Lett. \textbf{98}, 122506 (2011).

\bibitem{Vogt2009}K. Vogt, H. Schultheiss, S.~J. Hermsdoerfer , P. Pirro, A.~A. Serga, and B. Hillebrands, Appl. Phys. Lett. \textbf{95}, 182508 (2009).

\bibitem{Demidov2009b}V. E. Demidov, M. P. Kostylev, K. Rott, P. Krzysteczko, G. Reiss, and S.~O. Demokritov, Appl. Phys. Lett. \textbf{95}, 112509 (2009).

\bibitem{Slonczewski1996}J. Slonczewski, J. Magn. Magn. Mater. \textbf{159}, L1 (1996).

\bibitem{Demidov2011}V.~E. Demidov, S. Urazhdin, V.~V. Tiberkevich, A. Slavin, and S.~O. Demokritov, Phys. Rev. B \textbf{83} 060406 (2011).

\bibitem{Hermsdoerfer2009}S. Hermsdoerfer, H. Schultheiss, C. Rausch, S. Sch\"afer, B. Leven, S. Kim, and B. Hillebrands, Appl. Phys. Lett. \textbf{94}, 223510 (2009).

\bibitem{Bauer1997}M. Bauer, C. Mathieu, S.~O. Demokritov, and B. Hillebrands, J. Appl. Phys. \textbf{81}, 3971 (1997).

\bibitem{Demidov2009c}V.~E. Demidov, J. Jersch, K. Rott, P. Krzysteczko, G. Reiss, and S.~O. Demokritov, Phys. Rev. Lett. \textbf{102} 177207 (2009).

\bibitem{Bayer2004}C. Bayer, J.~P. Park, H. Wang, M. Yan, C.~E. Campbell, and P.~A. Crowell, Phys. Rev. B \textbf{69}, 134401 (2004).

\bibitem{Demidov2004}V.~E. Demidov, S.~O. Demokritov, B. Hillebrands, M. Laufenberg, and P.~P. Freitas, Appl. Phys. Lett. \textbf{85}, 2866 (2004).

\bibitem{NiFe}The material parameters used for calculating the spin-wave dispersions are:\newline
\mbox{Saturation magnetization $M_{s}=800~\textrm{kA/m}$}
\mbox{Gyromagnetic ratio: $\left|\gamma\right|/2\pi=28~\textrm{GHz/T}$}
\mbox{Exchange stiffness constant: $A=1.6\cdot10^{-11}~\textrm{J/m}$}

\bibitem{Donahue1999}M.~J. Donahue and D.~G. Porter, Interagency Report NISTIR 6376, National Institute of Standards and Technology, Gaithersburg, MD (1999).

\bibitem{Buttner1998}O. B\"uttner, M. Bauer, C. Mathieu, S.~O. Demokritov, B. Hillebrands, P. Kolodin, M. Kostylev, S. Sure, H. Dotsch, V. Grimalsky, Y. Rapoport, and A.~N. Slavin, IEEE Trans. Magn. \textbf{34}, 1381 (1998).

\bibitem{Demidov2008}V.~E. Demidov, S.~O. Demokritov, K. Rott, P. Krzysteczko, and G. Reiss, Phys. Rev. B \textbf{77}, 064406 (2008).

\bibitem{Jorzick2002}J. Jorzick, S. O. Demokritov, B. Hillebrands, M. Bailleul, C. Fermon, K. Y. Guslienko, A. N. Slavin, D. V. Berkov, and N. L. Gorn, Phys. Rev. Lett. \textbf{88}, 047204 (2002).

\end{thebibliography}
\end{document}